\documentclass{article}
\usepackage{PRIMEarxiv} 
\usepackage[utf8]{inputenc} % allow utf-8 input
\usepackage[T1]{fontenc}    % use 8-bit T1 fonts
\usepackage{hyperref}       % hyperlinks
\hypersetup{
colorlinks=true,
citecolor=green, 
urlcolor=blue
}
\usepackage{placeins}
\usepackage{float}
\usepackage{url}
\usepackage{siunitx}
\usepackage{booktabs}       % professional-quality tables
\usepackage{amsfonts}       % blackboard math symbols
\usepackage{nicefrac}       % compact symbols for 1/2, etc.
\usepackage{microtype}      % microtypography
\usepackage{lipsum}
\usepackage{fancyhdr}       % header
\usepackage{graphicx}       % graphics
\usepackage[shortlabels]{enumitem}
\usepackage[draft=true]{minted}
\setminted{
frame=lines,
framesep=2mm,
baselinestretch=1.2,
bgcolor=LightGray,
fontsize=\footnotesize,
linenos,
breaklines,
autogobble
}

\usepackage[dvipsnames]{xcolor}
\usepackage{color}
\usepackage{subfig}

\definecolor{LightGray}{gray}{0.95}
\graphicspath{{media/}} 
\usepackage{lscape}

\title{SAIPy: A Python Package for Single-Station Earthquake Monitoring using Deep Learning
}

\author{
    Wei Li\thanks{equal contribution} \\
    Frankfurt Institute for Advanced Studies, Germany \\
    \And
    Megha Chakraborty\footnotemark[1]\\
    Frankfurt Institute for Advanced Studies, Germany\\
    Institute of Geosciences, Goethe-University, Germany\\
    \And
    Claudia Quinteros Cartaya\\
    Frankfurt Institue for Advanced Studies, Germany\\
    \And
    Jonas Köhler\\
    Frankfurt Institute for Advanced Studies, Germany\\
    Institute of Geosciences, Goethe-University, Germany\\
    \And
    Johannes Faber\\
    Frankfurt Institue for Advanced Studies, Germany\\
    Institute for Theoretical Physics, Goethe Universität, Germany\\
    \And
    Georg Rümpker\\
    Frankfurt Institute for Advanced Studies, Germany\\
    Institute of Geosciences, Goethe-University, Germany\\
    \And
    Nishtha Srivastava\thanks{corresponding author, \href{mailto:srivastava@fias.uni-frankfurt.de}{srivastava@fias.uni-frankfurt.de}}\\
    Frankfurt Institute for Advanced Studies, Germany\\
    Institute of Geosciences, Goethe-University, Germany\\
}

\begin{document}
\maketitle

\begin{abstract}

Seismology has witnessed significant advancements in recent years with the application of deep learning methods to address a broad range of problems. These techniques have demonstrated their remarkable ability to effectively extract statistical properties from extensive datasets, surpassing the capabilities of traditional approaches to an extent. In this study, we present SAIPy, an open-source Python package specifically developed for fast data processing by implementing deep learning. SAIPy offers solutions for multiple seismological tasks, including earthquake detection, magnitude estimation, seismic phase picking, and polarity identification. We introduce upgraded versions of previously published models such as CREIME\_RT capable of identifying earthquakes with an accuracy above 99.8\% and a root mean squared error of 0.38 unit in magnitude estimation. These upgraded models outperform state-of-the-art approaches like the Vision Transformer network. SAIPy provides an API that simplifies the integration of these advanced models, including CREIME\_RT, DynaPicker\_v2, and PolarCAP, along with benchmark datasets. The package has the potential to be used for real-time earthquake monitoring to enable timely actions to mitigate the impact of seismic events. Ongoing development efforts aim to enhance SAIPy's performance and incorporate additional features that enhance exploration efforts, and it also would be interesting to approach the retraining of the whole package as a multi-task learning problem. A detailed description of all functions is available in a supplementary document.
\end{abstract}

% keywords can be removed
\keywords{Python package \and earthquake detection \and magnitude estimation \and seismic phase picking \and polarity estimation}

\section{Introduction}
Earthquake monitoring is an important step in the cataloging of earthquakes which records basic earthquake information such as start time, magnitude, location, etc. This forms the basis for understanding earth processes and analyzing potential seismic hazards \cite{li2021, li2021b}. Rapid earthquake monitoring involves a speedy detection of earthquakes followed by an estimation of important earthquake parameters such as phase arrival times, magnitude, etc. The detection of earthquakes has been traditionally performed by common algorithms such as short-time average/long-time average (STA/LTA) \cite{allen1978automatic}, which are heavily dependent on the choice of window lengths and thresholds and hence fail to perform satisfactorily over a wide range of signal-to-ratios (SNR). Seismic phase picking plays a vital role in many seismological workflows and is often performed manually. To achieve adequately automated seismic phase picking, many conventional approaches have been investigated over the past few decades. Notable algorithms developed for seismic phase picking include the STA/LTA \cite{allen1978automatic} and Akaike information criterion (AIC) \cite{takanami1991estimation}. However, these traditional methods fall short of achieving satisfactory performances when applied to low SNR signals. 

An earthquake's magnitude is considered to be one of ``the most important and also the most difficult parameters" \cite{jinetal} to estimate in real-time monitoring due to the complex nature of the geophysical processes that affect earthquakes \cite{Kanamori}. Consequently, even in the same magnitude scale, the magnitude reported by different agencies can vary by as much as 0.5 units \cite{Mousavi}. Traditionally frequency domain parameters such as predominant period $\tau_ p^{max}$ \cite{Nakamura} or amplitude domain parameters such as peak displacement ($P_d$) \cite{wuzhao} estimated from the first few seconds after P-arrival are used to estimate the magnitude using empirical relations. But these methods usually require distance information which is not always available in real-time. Another important parameter is the first-motion polarity for estimating focal mechanisms. It is usually assigned manually but aside from being time-consuming this is often riddled with human errors.

Deep learning, a data-driven technique, is widely applied in big data analysis tasks like image classification, speech recognition, and object detection \cite{lecun2015deep}. Unlike conventional machine learning approaches, deep learning algorithms prove their superiority in handling unstructured data such as images and text, and automating the process of feature extraction. As a consequence, it succeeds to reduce the dependency on human experts. Additionally, deep learning exhibits a remarkable capability to extract meaningful features from unlabeled data by decomposing the input data into hierarchical levels of representation. Besides, it also benefits from advanced computation technology like high-performing graphics processing units (GPUs). Over the past decades, the vast repositories of seismic data, coupled with the existence of hand-picked labels from earthquake catalogs, have opened up numerous possibilities for applying deep learning in seismology \cite{mousavi2022deep}. These applications include tasks such as seismic event detection, magnitude estimation, seismic phase picking,  and polarity estimation.

In this study, we introduce SAIPy, a Python package that combines several previously developed deep learning-based models such as CREIME \cite{chakraborty2022creime}, DynaPicker \cite{li2022dynapicker} and PolarCAP \cite{chakraborty2022polarcap}, and introduces an improved version of CREIME capable of processing data in real-time, CREIME\_RT and an improved version of DynaPicker, called DynaPicker\_v2. This package offers a unified interface, enabling users to conveniently access these models and two common datasets in seismology -- STEAD \cite{mousavi2019stanford} and INSTANCE \cite{michelini2021instance}. Additionally, users can retrain the models using their own seismic data, thereby enhancing their applicability and performance.

\section{SAIPy Framework}
In this section, we present an overview of the package's structure, along with the included models. Furthermore, we provide a pipeline to effectively integrate and utilize these models collectively.

\subsection{Data}
In the data blocks, we provide the interface to load and use two benchmark datasets-- 

$\bullet$ STanford EArthquake Dataset (STEAD ) \cite{mousavi2019stanford}: This is a high-quality global seismic dataset and contains local seismic signals including earthquake and non-earthquake traces. It includes about 1.2 million seismic waveforms, of which 200,000 are noise examples and the remaining contain seismic arrivals from 450,000 earthquakes with the magnitude $M$ ranging from -0.5 to 8. The waveforms are presented as NumPy arrays in units of counts and there are 35 and 8 metadata variables associated with each event and noise data respectively.

$\bullet$ INSTANCE dataset \cite{michelini2021instance}: It comprises 1.3 million regional 3-component waveforms, of which 50,000 are earthquakes with the magnitude $M$ ranging from 0 to 6.5, and 130,000 are noise examples. A total of 115 metadata variables are provided including a set of derived metadata like peak ground acceleration and velocity.

 The data block also contains functions to download and pre-process continuous seismograms from seismic stations (examples could be found in Section 3.1).

\subsection{Models}\label{models}
\begin{figure}[!htbp] 
    \begin{center}
        \includegraphics[width=\textwidth]{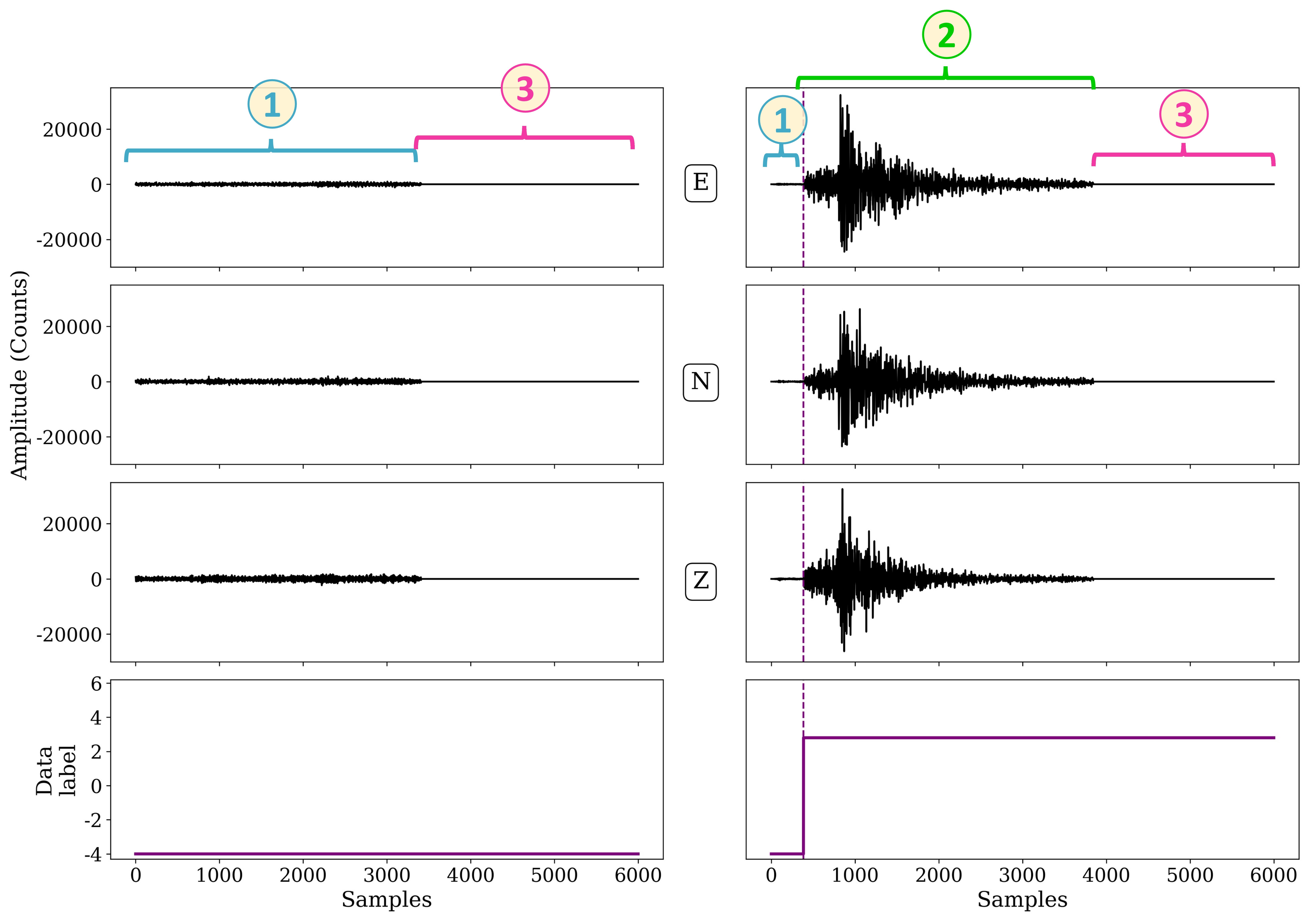}
        \caption{An example of data-labeling used to train CREIME\textunderscore RT. The different parts included in the waveforms are-- 1) (pre-signal) noise; 2) signal; and 3) zero-padding. The corresponding labels as described in Section 2.2 are shown in the bottom-most panel. Just like in CREIME \cite{chakraborty2022creime} an empirically tested value of -4 represents noise in the label, and following the P-arrival for events the label takes the value of the event magnitude.}
        \label{fig1}
    \end{center}
\end{figure}

\begin{figure}[!htbp] 
    \begin{center}
        \includegraphics[width=0.8\textwidth]{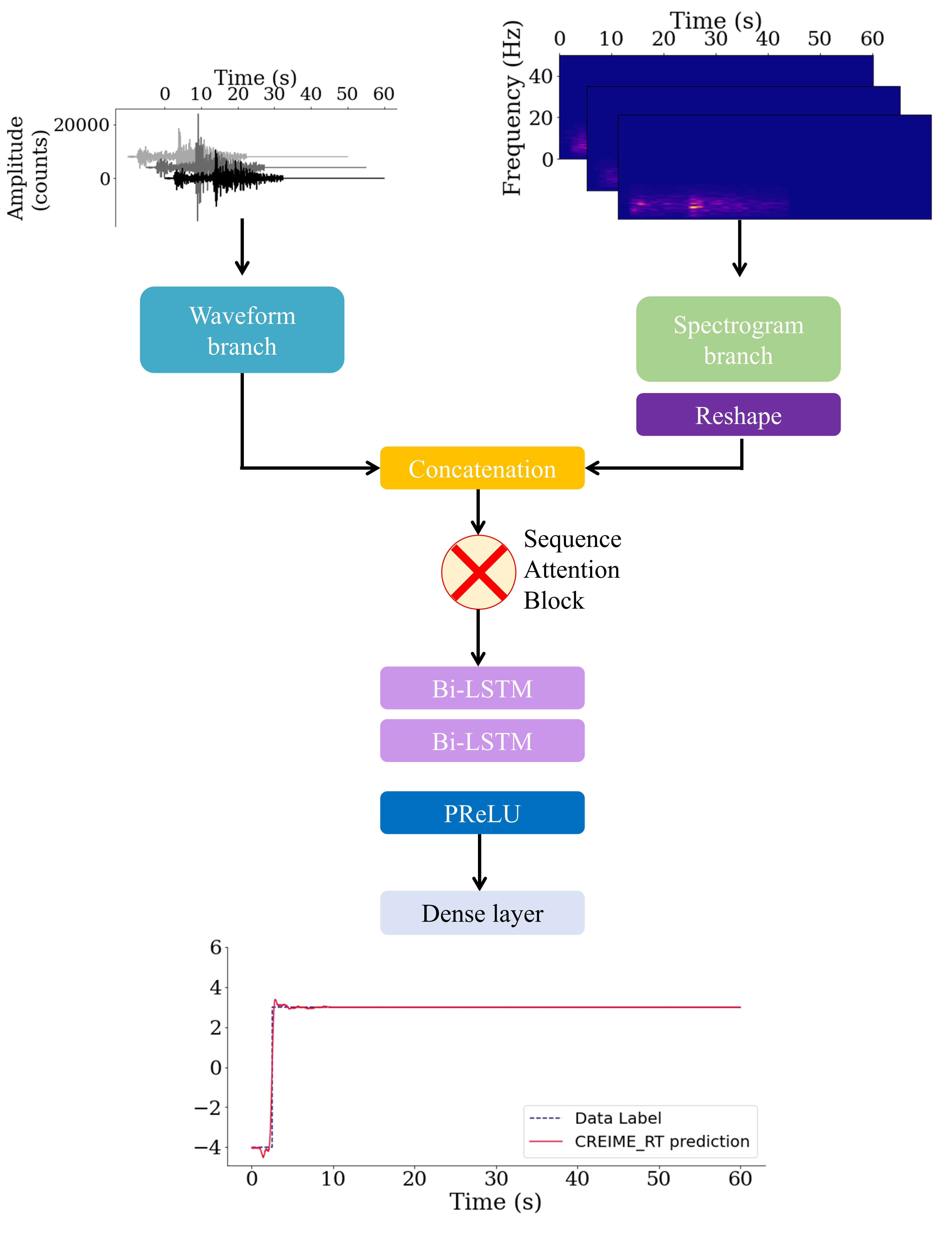}
        \caption{The architecture used for CREIME\_RT. We combine information from waveforms and their short-term Fourier transforms as input to the model using a concatenation layer. This is followed by a Sequence  Attention block (refer to the Transformer block in \cite{mousavi2020earthquake}) and Bidirectional LSTM layers. The final output is a dense layer of dimension 6000 $\times$ 1 corresponding to the 6000 input samples. For more details on the architecture please refer to the github page of the package.}
        \label{fig2}
    \end{center}
\end{figure}

\begin{figure}[!htbp] 
    \begin{center}
    \subfloat[]{
    \includegraphics[width=0.95\textwidth]{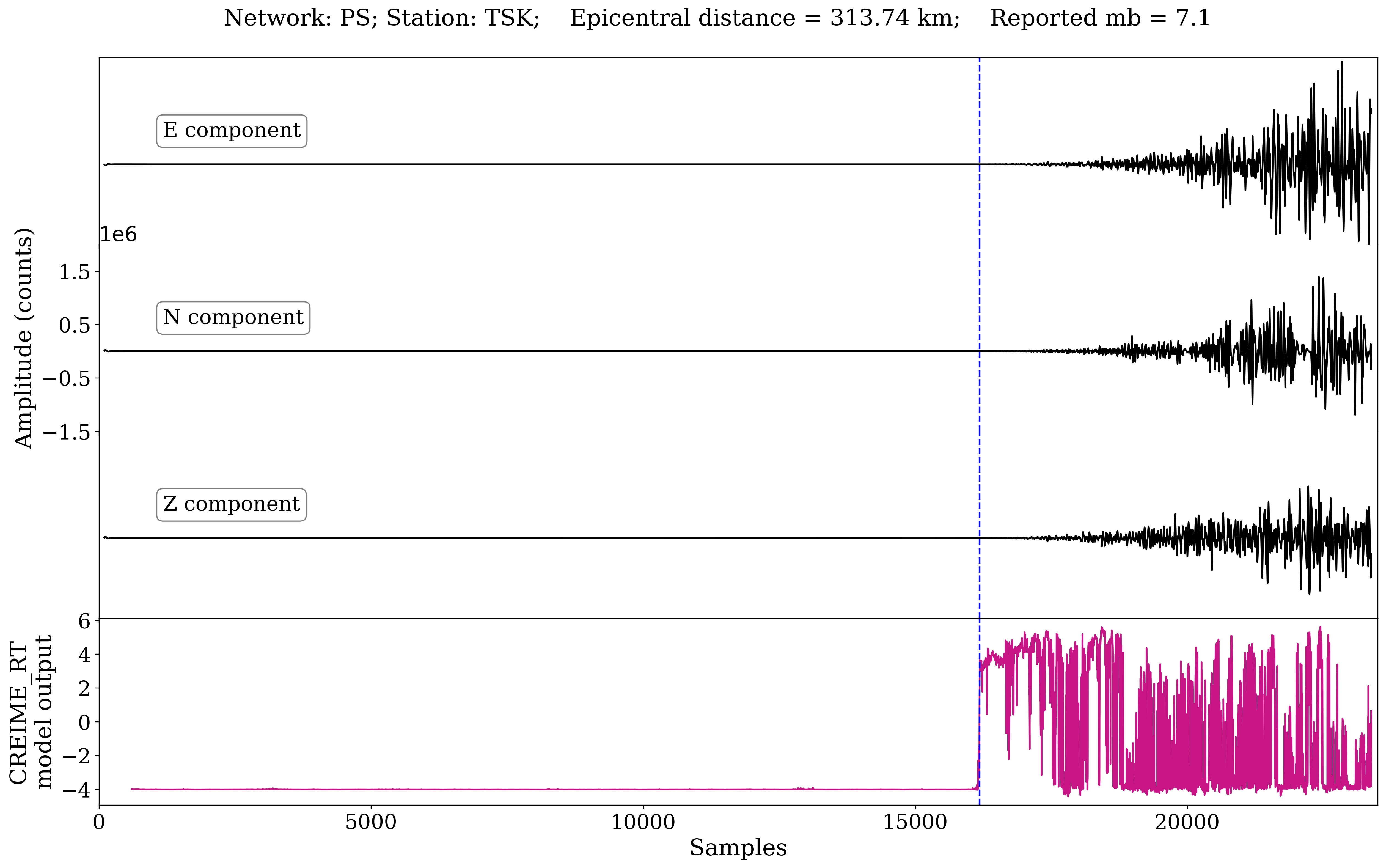}
    \label{fig31}}
    
    \subfloat[]{
    \includegraphics[width=0.95\textwidth]{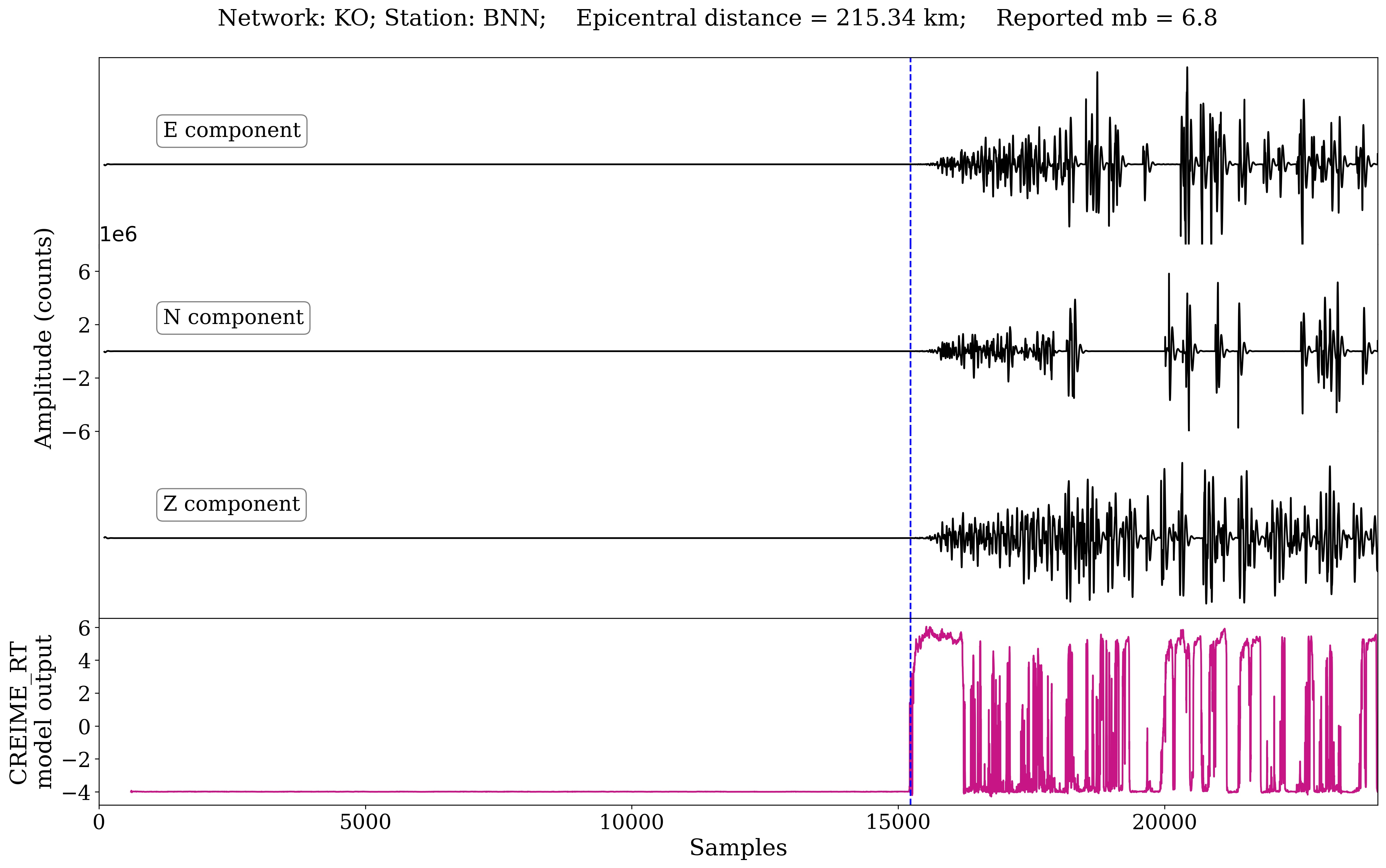}
    \label{fig32}}
1       \caption{Prediction of CREIME\_RT on a continuous waveform from (a) the 2011 Tohoku Earthquake (b) the 2023 Turkey Earthquake. The waveform (a) is downloaded from a station at Tsukuba, Japan (TSK) operated by Pacific21 (PS) network. The waveform (b) is downloaded from a station at Bunyan, Kayseri, Turkey (BNN) operated by the network Kandilli Observatory And Earthquake Research Institute (KO). Both waveforms are downloaded using Obspy \cite{beyreuther2010obspy}. The figure shows 3-component seismograms sampled at 100 Hz. The violet line shows the predictions by CREIME\_RT for continuous windows of length 5s shifted by 1 sample at a time. The blue line shows the P-arrival sample determined by DynaPicker \cite{li2022dynapicker} and used to test ViT model\cite{ViT}. Note that an output less than -0.5  is to be interpreted as noise. As one can observe there are gaps in the data for the Turkey earthquake which have been filled by zeros while downloading. The ViT \cite{ViT} model estimates magnitudes of 3.20 and 3.21 for these two waveforms respectively.}
    \label{fig3}
    \end{center}
\end{figure}

The model interface is an extensible framework that encompasses the application of our pre-developed models to seismic data. The pre-developed models on the data at a sampling rate of 100 $\unit{Hz}$ integrated into the package are listed below\footnote{Interested readers are encouraged to read proper references for further details about each model.}. It should be noted, the PolarCAP model utilizes only one component data, whereas all other models necessitate three-component data. 
\begin{itemize}[leftmargin=*]
    \item CREIME \cite{chakraborty2022creime}, a Convolutional Recurrent model for Earthquake Identification and Magnitude Estimation, which can simultaneously perform earthquake identification, local magnitude estimation, and first P-wave arrival time regression solely based on 1–2~s P-wave recording preceded by pre-signal noise.
    \item CREIME\_RT, which uses the basic idea of  CREIME and is capable of processing data of variable length which makes it easier to implement in real-time. This is achieved with the help of zero-padding as shown in Figure \ref{fig1}; the total length of data used is always 60 s (or 6000 samples). This model employs short-term Fourier transforms (STFT) of the data and sequence attention blocks to extract relevant information from the data; a schematic of its architecture can be found in Figure \ref{fig2}. The original distribution of magnitudes in the training dataset follows the Gutenberg-Richer law \cite{gutenberg1944frequency} which can be very detrimental to the model learning \cite{imbalance}. So we randomly undersample the events with magnitude up to 3.5 and randomly oversample events with magnitude above 4.0. To further improve the prediction of higher magnitudes, additional high-magnitude events ($m_L \geq 5.5$) are hand-picked, manually downloaded and added to the training data. For more details on the model training please refer to the Supplementary Information. The model achieves an accuracy of 99.80\% in signal-vs-noise discrimination on a test dataset from STEAD \cite{mousavi2019stanford} and can estimate magnitudes with a root mean squared error (RMSE) of 0.38. The RMSE for events with a magnitude above 4 is 0.55. We compare the performance of CREIME\_RT with the vision transformer (ViT) network \cite{ViT} on the INSTANCE dataset \cite{michelini2021instance} and observe that the RMSE in the prediction of all magnitudes by CREIME\textunderscore RT and ViT are 0.62 and 0.53, respectively and for magnitudes higher than 4 the RMSE are 0.97 and 1.30 respectively. Thus, CREIME\_RT performs better for higher magnitudes as compared to ViT. We also test the performance of CREIME\_RT on continuous data from 4 notorious earthquakes-- the 2011 Tohoku Earthquake ($M_w$ 9.1), the 2023 Turkey Earthquake ($M_w$ 7.8), the 2015 Nepal Earthquake ($M_w$ 7.8), and the 2014 Iquique Earthquake ($M_w$ 8.2) and find that CREIME\_RT satisfactorily detects and estimates the magnitudes for all these Earthquakes. We further observe that the ViT model considerably under-estimates the magnitude for each of these large events (to get the correct window for the application of ViT to these waveforms we use the P-arrival time picked by DynaPicker \cite{li2022dynapicker}). The example for the 2011 Tohoku Earthquake and the 2023 Turkey Earthquake is shown in Figure \ref{fig3}. One can visualize the 3-component waveforms from a single station and the magnitude estimated by CREIME\_RT for shifting data windows. As one can see from Figure \ref{fig1}, a value of -4 represents noise. For the other two examples please refer to the Supplementary Information.
    
    \item DynaPicker\_v2, an improved version of the previously developed DynaPicker \cite{li2022dynapicker}, is a deep learning-based solution for seismic body wave phase classification and picking. Different from the original DynaPicker \cite{li2022dynapicker}, which is built on dynamic convolution \cite{li2021revisiting}, DynaPicker\_v2 mainly employs a large kernel size and replaces the first normal convolution layer with the 1D dynamic convolution decomposition (DCD) block adapted from the image classification case \cite{li2021revisiting}. Furthermore, the activation function is also updated\footnote{More details can be found in the package code part.}. The activation function used in each 1D-DCD layer is Gaussian error linear units (GELU) \cite{hendrycks2016gaussian}, and batch normalization\cite{ioffe2015batch} is applied to each 1D-DCD layer. The supplementary information includes the confusion matrices for DynaPicker and DynaPicker\_v2, revealing that DynaPicker\_v2 outperforms in phase classification. It is pre-trained on the short-window format data (e.g., 4s dataset from the Southern California Earthquake Data Center (SCEDC) \cite{center2013southern}), and is further used to estimate the arrival times of the P- and S-waves on the continuous waveform on a long-time scale.
    
    \item PolarCAP \cite{chakraborty2022polarcap} is developed on the autoencoder architecture that aims to identify first-motion polarities of earthquake waveforms. PolarCAP is trained in a supervised fashion using labeled traces from INSTANCE \cite{michelini2021instance} dataset and is cross-validated to choose the most optimal set of hyperparameters. It is demonstrated to be more consistent and time-saving than manual picking of polarities.
\end{itemize}

\newpage

\subsection{Practical application: a combined pipeline for real-time data}
\begin{figure}[!htbp]
    \centering
    \includegraphics[width=0.78\textwidth]{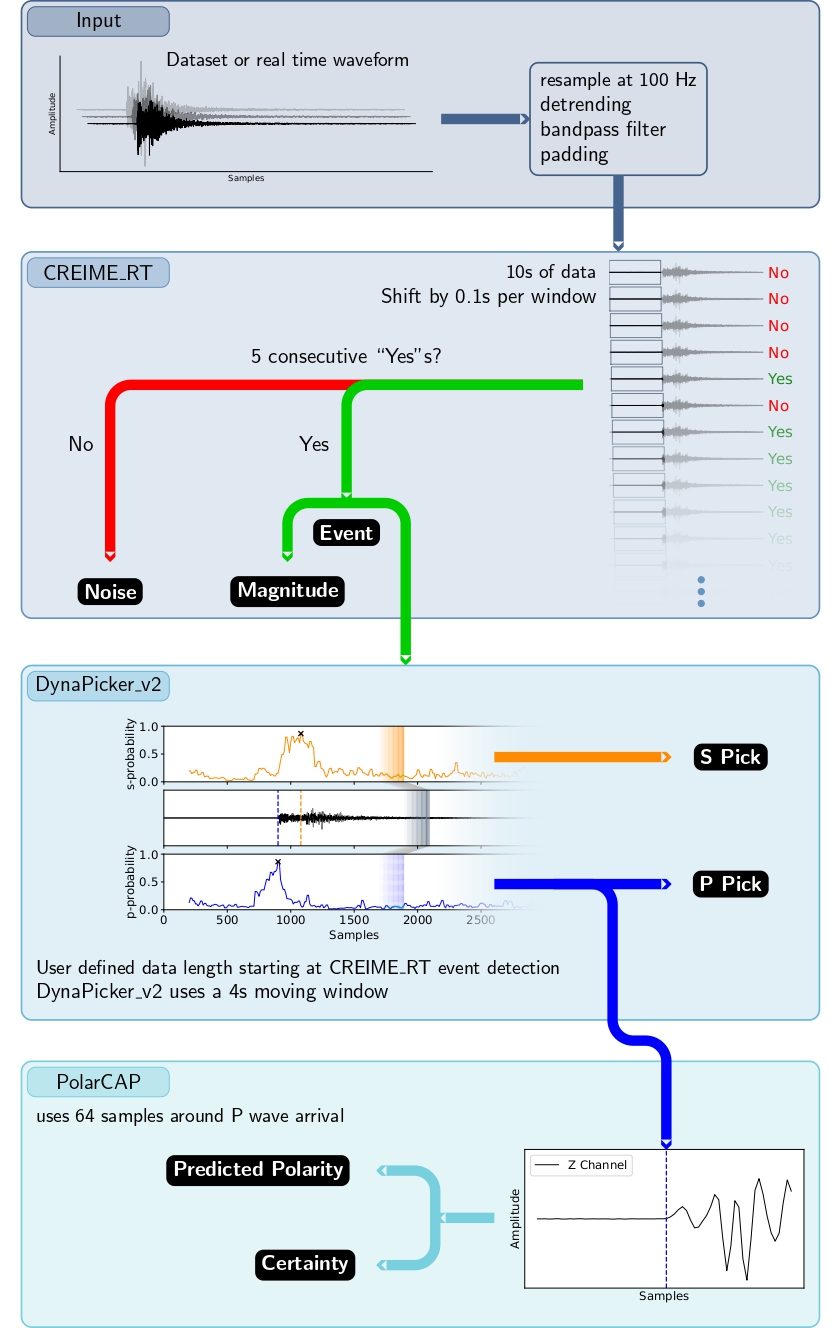}
    \caption{Schematic diagram for SAIPy. The input (continuous waveform or sample from a dataset) is resampled and preprocessed. Then CREIME\_RT uses 10~s windows shifted by 0.1~s to detect earthquake signals. Once five consecutive windows are found to contain an earthquake signal, the magnitude will be determined and an event trigger is placed. As long as there are no 5 consecutive windows found to contain earthquakes, the trace is considered noise.
    An earthquake signal is processed with DynaPicker\_v2 in the next step. There, moving 4~s windows are used to determine the P- and S- wave arrivals. The end of the 4~s moving window is shown by the vertical line in the middle panel of the DynaPicker\_v2 box. 
    As a final step, the P-pick is used to select the Z-Channel data of 31 samples before and 32 samples after, which is then fed to PolarCAP for polarity prediction.
    All model outputs are marked in black boxes.}
    \label{fig4}
\end{figure}

As shown in Figure \ref{fig4}, the workflow is summarized as follows.
% \begin{enumerate}[(a),leftmargin=*]
\begin{itemize}[leftmargin=*]
    \item First, given continuous seismic waveform, the CREIME\_RT model is adopted for event detection\footnote{It is worth noting that in this pipeline, CREIME\_RT is employed prior to the picker model with the goal of minimizing the chances of misclassifying phase picking.}. For this, sliding windows of length 10~s (with zero-padding), shifted by 10 samples at a time are fed to CREIME\_RT. Each window is classified into event or noise.
    \item When 5 consecutive windows are detected as an event, the waveform captured from the beginning of the first of these windows and up to a user-defined length is fed into DynaPicker\_v2 for phase picking. 
    \item Magnitude estimation is performed by feeding a data window of length 6000 samples of data (with zero-padding if there is not enough data) starting from the first sample of the first of the five consecutive windows used for detection to CREIME\_RT.
    \item Finally, 64 sample windows of the raw waveform centered around the estimated P-phase sample are used as input into the PolarCAP model to determine its polarity.
% \end{enumerate}
\end{itemize}

\begin{figure*}[!htbp] 
    \subfloat[]{}
    \includegraphics[width=0.9\textwidth]{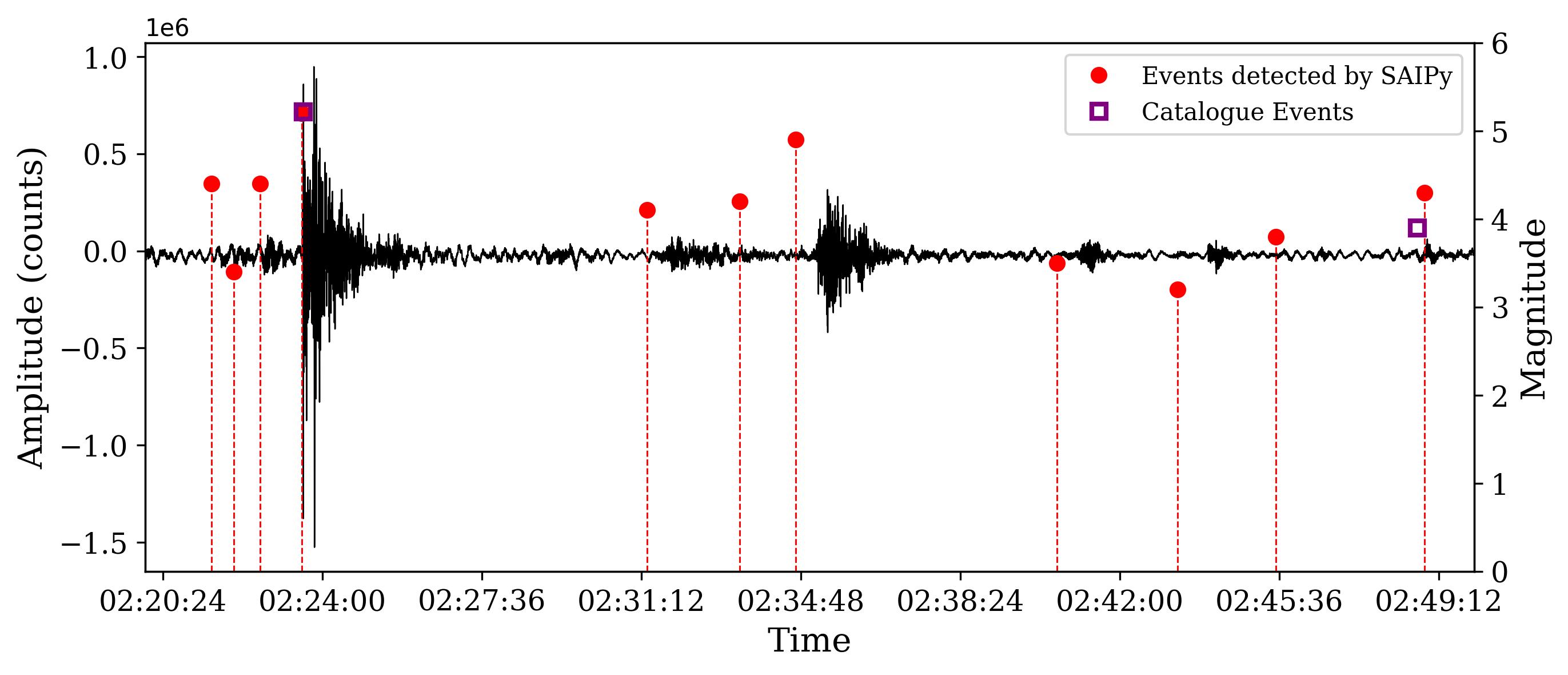}
    
    \subfloat[]{}
        \includegraphics[width = 0.45\textwidth]{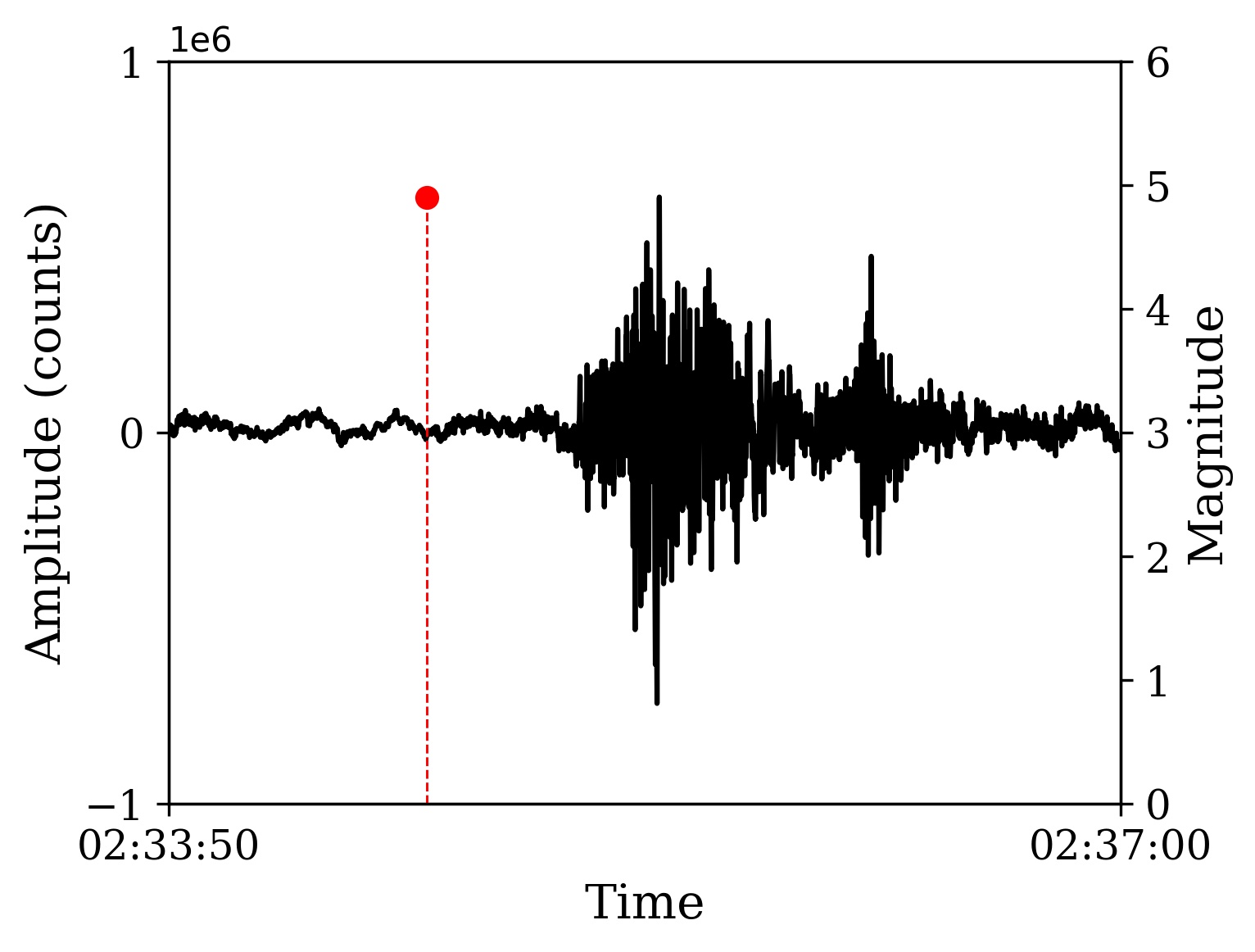}
    \subfloat[]{}
        \includegraphics[width = 0.45\textwidth]{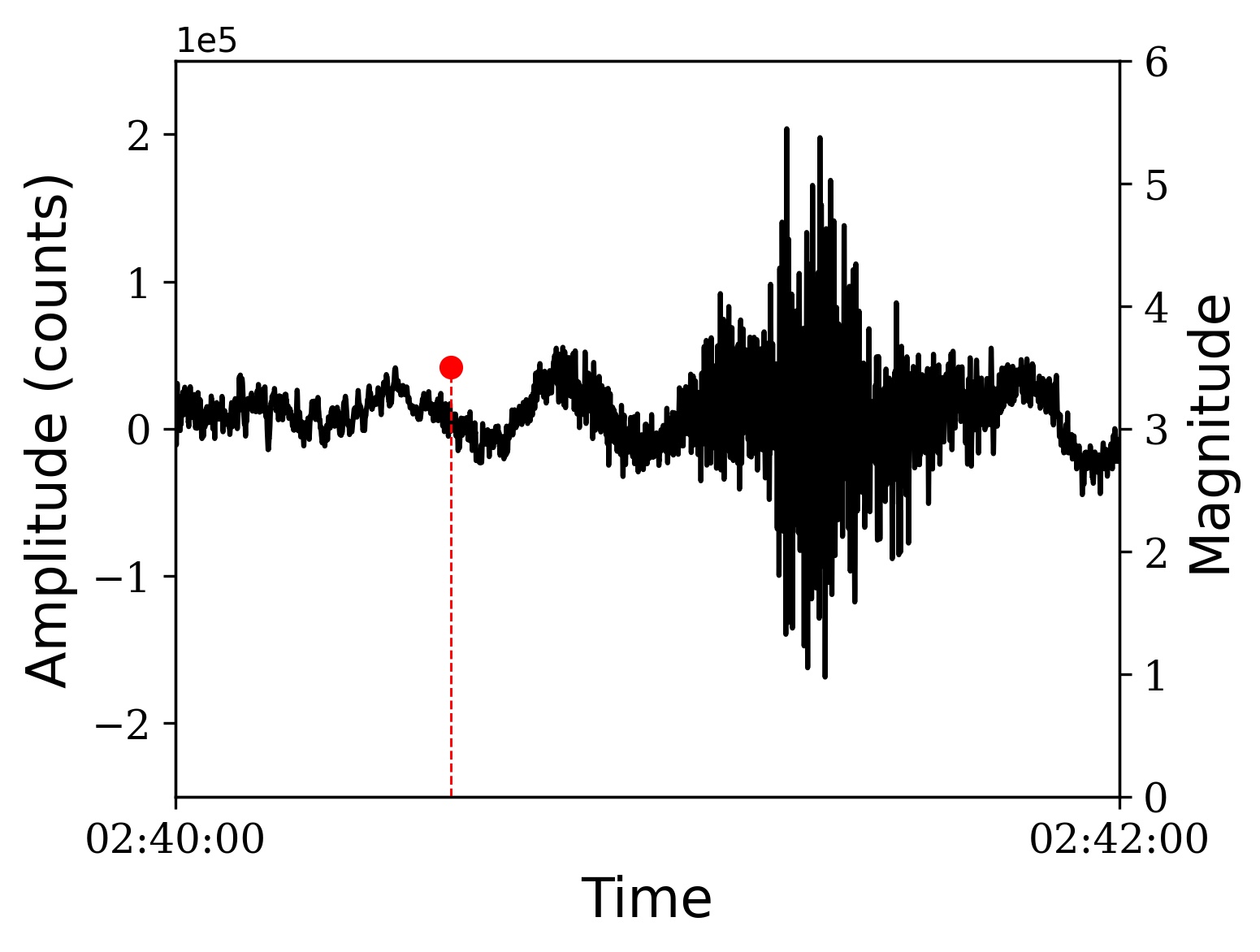}
   \caption{Predictions of the SAIPy package on one of the aftershock sequences of the 2023 Turkey earthquake downloaded from the station CEYT (Ceyhan, Adana, Turkey) belonging to the Kandilli Observatory And Earthquake Research Institute (KOERI) network. As one can see from the figure SAIPy detects and provides a satisfactory magnitude estimate for the events present in the catalogue. Upon zooming in we notice that SAIPy also seems to detect uncatalogued events.}
    \label{fig5}
    % \end{center}
\end{figure*}

In this paper, we present an example of the application of the SAIPy package to the aftershock sequences of the 2023 Turkey earthquake recorded at CEYT (Ceyhan, Adana, Turkey) station belonging to the Kandilli Observatory And Earthquake
Research Institute (KOERI) network (Figure \ref{fig5}). SAIPy not only detects and satisfactorily estimates the magnitude of the events in the catalogue, but also detects events that are not catalogued. The zoomed in plots show that these detections correspond to what appears to be seismic events.

% \begin{figure}[t]
%     \centering
%     \includegraphics[width=0.8\textwidth]{Figures/SAIPy.jpg}
%     \caption{Schematic diagram for SAIPy. The start 4~s moving window of DynaPicker\_v2 is shown by the vertical line in the middle panel of the DynaPicker\_v2 box.}
%     \label{fig4}
% \end{figure}

\section{Example workflows}
In this section, we emphasize the process of data loading, the utilization of pre-trained models for testing purposes, and the methodology involved in retraining those models\footnote{Kindly ensure that you substitute the placeholder `data\_path' with the actual path where you intend to store your data when utilizing the provided code snippets.}.

\subsection{Workflow 1: Loading data}
The package data module contains functionality to read and process seismological datasets. Here, we give a tutorial on how to use two existing datasets -- the STanford EArthquake Dataset (STEAD) \cite{mousavi2019stanford} and the INSTANCE benchmark dataset \cite{michelini2021instance}, and how to download older data for analysis. The commands to do this are listed below.

\subsubsection{Benchmark data loading}
For these two common datasets, the data loading processes are shown as follows:
\begin{minted}{python}
from saipy.data.base import STEAD, INSTANCE
stead, instance = STEAD('data_path'), INSTANCE('data_path')
\end{minted}

\subsubsection{Continuous data downloading}
To download continuous data\footnote{Both the DynaPicker\_v2 model and the package have the capability to directly utilize continuous data.}, we utilize the ``get\underline{\space}waveforms'' function provided by the ObsPy package \cite{beyreuther2010obspy}. This function allows to retrieve waveform data for a specified network, station, location, and channel within a specified time duration from the start time to the end time. The process of downloading the continuous data is illustrated as follows, and it is equivalent to the ObsPy package.
\begin{minted}{python}
from saipy.data.realdata import *
# both starttime and endtime are in UTCDateTime format
stream = waveform_download(wsp, net, sta, loc, chan, starttime, endtime)
\end{minted}

It must be noted that in certain situations, the data from the client or network may not be accessible resulting in an error message like "No data available for request." In this case, it is necessary to attempt alternative clients in order to resolve the issue.

\subsection{Workflow 2: Using pre-trained models}
The following code blocks show how to load each of the pre-trained deep learning models described in Section \ref{models} and apply the model to data provided in the correct input format.
\subsubsection{Using pre-trained CREIME}
The following steps are needed to use the pre-trained CREIME and using it on STEAD dataset \cite{mousavi2019stanford}.
\begin{minted}{python}
# loading CREIME from package
from saipy.models.creime import CREIME
creime = CREIME()

# loading STEAD from package
from saipy.data.base import STEAD
stead = STEAD(directory='data_path')

# getting data in CREIME format
X, y = stead.get_creime_data(stead.trace_list())

# making predictions
y_pred, predictions = creime.predict(X)
\end{minted}

\subsubsection{Using pre-trained CREIME\textunderscore RT}
The following steps are needed to use the pre-trained CREIME\textunderscore RT on the STEAD dataset \cite{mousavi2019stanford}.

\begin{minted}{python}
# loading CREIME_RT from the package
from saipy.models.creime import CREIME_RT
creime_rt = CREIME_RT()

# loading STEAD from package
from saipy.data.base import STEAD
stead = STEAD(directory='data_path')

# getting data in CREIME_RT format
X, y = stead.get_creime_rt_data(stead.trace_list())

# making predictions
y_pred, predictions = creime_rt.predict(X)
\end{minted}

\subsubsection{Using pre-trained DynaPicker\_v2} 
Below is an example demonstrating the utilization of the pre-trained DynaPicker\_v2 model on the STEAD dataset \cite{mousavi2019stanford}.

\begin{minted}{python}
from saipy.data.base import STEAD
from saipy.models.dynapicker import load_model

# load the model from default path
model_path = ''
model = load_model(model_path)

# device information
device = torch.device('cuda' if torch.cuda.is_available() else 'cpu')
print("Device: ", device)

# loading data 
stead = STEAD('data_path')
metadata,waveform = stead.get_dynapicker_data()

# trace selection
df = metadata[(metadata.trace_category == 'earthquake_local')]
ev_list = df['trace_name'].to_list()[10:11]
dataset = waveform.get('data/'+str(ev_list[0])) 

# hdf5 dataset into obspy.stream
stream = make_stream_stead(dataset) 

# phase picking 
prob_p, prob_s, pwave, swave = phase_picking(device, model, stream, bandpass_filter_flag=True, picker_num_shift=10, batch_size=4, fremin=1, fremax=40, fo=5, fs=100)
\end{minted}

\subsubsection{Using pre-trained PolarCAP}
Below are the steps to use pre-trained PolarCAP on a chunk of the STEAD dataset \cite{mousavi2019stanford}.
\begin{minted}{python}
#loading PolarCAP from package
from saipy.models.polarcap import PolarCAP
polarcap = PolarCAP()

# loading STEAD from package
from saipy.data.base import STEAD
stead = STEAD(directory='data_path')

# getting data in the correct format for PolarCAP
X = stead.get_polarcap_data(stead.trace_list())

# making predictions
predictions = polarcap.predict(X)
\end{minted}

\subsection{Workflow 3: Using pipeline for continuous data}
Here we show how to use the package to monitor continuous data.

\begin{minted}{python}
# specifying waveform details
t1 = UTCDateTime("2021-07-31T21:05:00")
wsp = 'IRIS'
network = 'AE'
station = '319A'
location = '*'
channel = "*"
start_time = t1
end_time = t1+60*60

# monitoring continuous data
device = torch.device('cuda' if torch.cuda.is_available() else 'cpu')
result_dict = earthquake_monitor(wsp, network, station, location, channel, start_time, end_time, device, leng_win=120, save_result=True, path='', file_name='result.csv')
\end{minted}

\subsection{Workflow 4: Retraining individual models}
The following code blocks show how to retrain the models listed in this study on new data. It is also possible to employ transfer learning by loading the trained model (by setting the `untrained' parameter to False), and then training it further on a different dataset.

\subsubsection{CREIME}
The following code block shows how to retrain CREIME on the data from STEAD dataset \cite{mousavi2019stanford}.

\begin{minted}{python}
# loading the untrained CREIME model
from saipy.models.creime import CREIME
creime = CREIME()
model = creime.get_model(untrained=True)

# loading training data from STEAD dataset 
from saipy.data.base import STEAD
stead = STEAD(directory='data_path')
X, y = stead.get_creime_data(stead.trace_list())

# retraining model 
model.fit(X, y, epochs = 10)
\end{minted}

\subsubsection{CREIME\textunderscore RT}
The following steps can be used to retrain CREIME\textunderscore RT using the STEAD dataset \cite{mousavi2019stanford}.
\begin{minted}{python}
# loading the untrained model
from saipy.models.creime import CREIME_RT
creime_rt = CREIME_RT()
model = creime_rt.get_model(untrained=True)

# loading training data from STEAD
from saipy.data.base import STEAD
stead = STEAD(directory='data_path')
X, y = stead.get_creime_rt_data(stead.trace_list(), training=True)

# retraining model
model.fit(X,y,epochs=10)
\end{minted}
\subsubsection{DynaPicker\textunderscore v2} In order to retrain DynaPicker\_v2, the following main steps are used.

\textbf{Step 1: setting up.}
\begin{minted}{python}
from saipy.modules.phaseclassification import *
from saipy.models.dynapicker import *
# reproduce
set_seed(24) 
args = arguments()
# loss function
criterion = torch.nn.CrossEntropyLoss()
# optimizer
optimizer = torch.optim.Adam(model.parameters(), lr=args.lr)
\end{minted} 

During this stage, users can conveniently configure hyperparameters such as learning rate and batch size by assigning new values to the respective variables through the use of "args.variable = new value".

\textbf{Step 2: picker retraining.}
\begin{minted}{python}
train_dataset, valid_dataset = CustomDataset(train_x, train_y), CustomDataset(valid_x, valid_y)
# dataloder
Train_Loader = torch.utils.data.DataLoader(dataset=train_dataset, batch_size=args.batch_size, shuffle=True, num_workers=0)
Valid_Loader = torch.utils.data.DataLoader(dataset=valid_dataset, batch_size=args.batch_size, shuffle=False, num_workers=0)
retrained_model, train_loss, valid_loss = train(args, device, Train_Loader, Valid_Loader, criterion, optimizer, scheduler=None)
\end{minted}
        
\textbf{Step 3: picker testing.}
\begin{minted}{python}
test_dataset = CustomDataset(test_x, test_y) 
Test_Loader = torch.utils.data.DataLoader(dataset=test_dataset, batch_size=args.batch_size, shuffle=False)
preds, pred_prob = test(args, device, retrained_model, Test_Loader, criterion)
\end{minted}

For more elaborate examples of plotting data, training history, confusion matrix, metrics report, precision-recall curve, and ROC curve, please refer to the example files provided within the corresponding package. These files contain more comprehensive details and instructions on how to plot these visualizations.

\subsubsection{PolarCAP}
The following steps can be used to retrain PolarCAP using the INSTANCE dataset \cite{michelini2021instance}

\begin{minted}{python}
# loading untrained PolarCAP model
from saipy.models.polarcap import PolarCAP
polarcap = PolarCAP()
model = polarcap.get_model(untrained=True)

# loading INSTANCE dataset for training
from saipy.data.base import INSTANCE
instance = INSTANCE('/home/seismoai/data')
X, y = instance.get_polarcap_data(instance.trace_list_events())

# retraining model
model.fit(X, y, epochs=10)
\end{minted}

\section{Discussion: Continual improvement process}

We have introduced the SAIPy package which is capable of automatically monitoring continuous data. The SAIPy package has been demonstrated to be a useful tool to analyze the large volumes of data that have been archived over the years from permanent and temporary networks for further analysis and research. Furthermore, we are endeavoring to amplify its capabilities for real-time monitoring implementations.

The continuous collection, processing, and analysis of seismic data to detect, locate, and characterize earthquakes in real-time monitoring typically involves several components working together to process large amounts of seismic data quickly, enabling decision-making and significantly enhance emergency response systems.

SAIPy package has been trained to automatically extract relevant features from raw 3-component data streams from each station independently, providing earthquake detection, magnitude estimation, phase picking, and first-motion polarity identification. Subsequently, the information for each station provided by SAIPy can be jointly used for other algorithms to estimate the epicenter, study the focal mechanism and generate automated reports.

All the models included in the package are pre-trained on the data at a sampling rate of 100 $\unit{Hz}$. So far, we are uncertain of the picking performance on the data sampled at a different rate. Also, the seismic data used for the pre-training process for DynaPicker\_v2 has been limited to epicentral distances below 100 $\unit{km}$. 

Additionally, the magnitude estimation for large earthquakes is a challenge in many cases, since the data recorded from inertial sensors, such as seismometers, tend to be clipped when these sensors are at a short distance from the epicenter. As the models included in SAIPy work using exclusively seismograms, the package has optimal performance primarily ensured for small to moderate earthquakes. While the data augmentation and training techniques used for the CREIME\_RT model is able to minimise under-estimation of magnitudes upto 5.5, some under-estimation is still observed for higher magnitudes. This leaves us with the task to continuously look for architectures and methods to improve model performance for such magnitudes. We aim to address and alleviate these constraint in short-term future works in order to enable the application of SAIPy not only to real-time monitoring but also to contribute to early warning systems.

We have consolidated all the models for direct applications. Additionally, each individual model can be retrained separately. These deep learning models have the ability to adapt and generalize to new data. As more earthquake data becomes available, the models can be continuously retrained to improve their performance and accuracy. This adaptability allows the models to learn from diverse seismic events and adapt to different regions or geological contexts, making them versatile tools for earthquake monitoring in various locations around the world. The prospect of retraining the entire package in the future presents an intriguing avenue to explore. For instance, it would be interesting to approach the retraining of the whole package as a multi-task learning problem \cite{ruder2017overview}. 

Another aspect that deserves attention is the inherent presence of noise content in the data. Numerous conventional techniques have been devised and employed to improve the signal-to-noise ratio (SNR) of these recorded signals. However, these methods prove ineffective when the frequencies of the noise and the seismic data are overlapped. Choosing the most suitable filtering parameters is also a challenging task as they tend to change over time and can significantly impact the waveform's shape, consequently diminishing subsequent analysis quality \cite{zhu2019seismic}.

Therefore, numerous denoising solutions based on deep learning have been developed demonstrating their remarkable effectiveness in seismic denoising as compared to conventional approaches \cite{mousavi2022deep}. These solutions can be broadly categorized into two main categories: supervised \cite{zhu2019seismic,yang2022toward} and unsupervised approaches \cite{saad2022unsupervised}. Supervised denoising methods necessitate a substantial amount of clean and noisy data pairs in order to achieve optimal results. On the other hand, there are instances where the unsupervised method proved ineffective in denoising data e.g., some noise components persist or the P-wave samples of the denoised data are missed \cite{saad2022unsupervised}.

There are still several challenges faced by unsupervised seismic denoising methods. Hence, we will persist in our efforts to improve seismic denoising, and then incorporate the denoising process into our package. For future enhancements, the package could also incorporate additional models, such as an algorithm for source-distance estimation, thereby expanding the versatility of SAIPy. 

\section{Conclusion}
We have developed SAIPy as an open-source Python package, built to aid users in their application of deep learning techniques to seismic data. It minimizes common barriers to development for users looking to apply deep learning methods to seismic tasks (e.g., earthquake detection, magnitude estimation, body-wave phase picking, and polarity estimation). We introduce a pipeline to combine several models to ensure a smooth workflow for monitoring continuous data whereby events are automatically detected and their P- and S-arrival times, magnitudes, and first-motion polarity are characterized.  The scalability of these deep learning models is essential for monitoring earthquake activity over vast regions and handling the increasing volume of data generated by seismic networks. Hence, we emphasize that the codes will undergo regular updates to ensure its continued optimal functionality.

\section*{Code availability}
All codes in this work are available at \url{https://github.com/srivastavaresearchgroup/SAIPy}. In this package, both the CREIME\_RT and PolarCAP models are implemented in Keras \cite{chollet2015keras}, and the Dynapicker\_v2 model is implemented in PyTorch \cite{paszke2019pytorch}. 

\section*{Acknowledgments}
This research is supported by the ``KI-Nachwuchswissenschaftlerinnen'' - grant SAI 01IS20059 by the Bundesministerium für Bildung und Forschung - BMBF. Calculations were performed at the Frankfurt Institute for Advanced Studies' GPU cluster, funded by BMBF for the project Seismologie und Artifizielle Intelligenz (SAI). 

\bibliographystyle{unsrt}
\bibliography{reference}  
\end{document}